\documentclass[fleqn,10pt]{wlscirep}

\title{Crumpling Damaged Graphene}

\author[1,*]{I.Giordanelli}
\author[1]{M. Mendoza}
\author[1,2]{J. S. Andrade, Jr.}
\author[3]{M. A. F. Gomes}
\author[1,2]{H. J. Herrmann}

\affil[1]{ETH
  Z\"urich, Computational Physics for Engineering Materials, Institute
  for Building Materials, Wolfgang-Pauli-Strasse 27, HIT, CH-8093 Z\"urich, Switzerland}
\affil[2]{Universidade
  Federal do Cear\'a, Departamento de F\'isica, Campus do Pici, 60455-760 Fortaleza, Cear\'a, Brazil}
\affil[3]{Universidade Federal de Pernambuco, Departamento de F\'isica, 50670-901 Recife-PE, Brazil}

\affil[*]{gilario@ethz.ch}

\begin{abstract}
Through molecular mechanics we find that non-covalent interactions modify the fractality of crumpled damaged graphene.
Pristine graphene membranes are damaged by adding random vacancies and carbon-hydrogen bonds. Crumpled membranes exhibit a fractal dimension of $ 2.71 \pm 0.02$ when all interactions between carbon atoms are considered, and $2.30 \pm 0.05$ when non-covalent interactions are suppressed. The transition between these two values, obtained by switching on/off the non-covalent interactions of equilibrium configurations, is shown to be  reversible and  independent on  thermalisation. In order to explain this transition, we propose a theoretical model that is compatible with our numerical findings. Finally, we also compare damaged graphene membranes with other crumpled structures, as for instance, polymerised membranes and paper sheets, that share similar scaling properties.
\end{abstract}


\begin{document}
\flushbottom 
\maketitle

\thispagestyle{empty}

\section*{Introduction}

Graphene is   a one-atom thick membrane  possessing  extraordinary mechanical and electronic properties \cite{Novoselov26072005,Novoselov22102004,Lee18072008}. It naturally forms ripples even at zero temperature, overcoming the restriction on long-range $2$d order imposed by the Mermin-Wagner theorem \cite{intrinsic-ripples}. Graphene sheets can be systematically damaged, for instance by creating vacancies through irradiation \cite{Kotakoski2011}  and, in some specific cases, they can transform into other carbon structures. \cite{chuvilin2010direct} Remaining unsaturated carbon bonds are quite reactive but can be neutralized with hydrogen. 
 The initially flat neutralized hydro-carbon structure, which we call damaged graphene membrane (DGM), has the tendency to crumple. 
How this crumpling takes place and on which ingredients it depends, is the subject of the present Report.

\begin{figure}
\begin{center}
\includegraphics[width=0.5\columnwidth]{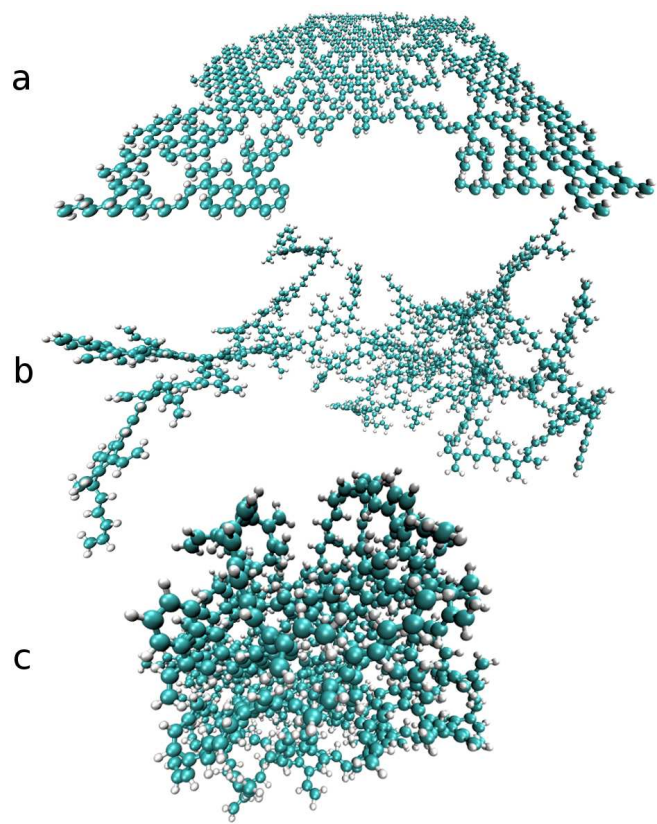}
\caption{Typical realization of a damaged graphene membrane (DGM) generated with a vacancy density of $p_c=0.303$ and a system size of $L=60$ \AA.  
Green spheres represent  carbon and  white spheres hydrogen atoms. 
The initial configuration is shown in (a). The corresponding crumpled DGM after  $1.8$ ns at
$T=0$ K is shown in (b) without NCIs, and in (c) with NCIs.}
\label{fig:initial_final_p69}
\end{center}
\end{figure}

Through sophisticated molecular mechanics simulations we show here that, for sufficient degree of damage, graphene sheets become fractal with a dimension that depends on the interaction range of the potential. Non-covalent interactions (NCIs) like van der Waals forces play such a dominant role,
that they not only densify the molecular structure but even enhance the fractal dimension of DGM by over $20\%$.

\section*{Method}
 
We simulate DGMs using molecular mechanics with the adaptive intermolecular reactive bond-order (AIREBO) potential \cite{stuart2000reactive}. This many-body potential has been developed to simulate molecules of  carbon and hydrogen. The AIREBO potential is defined by a sum over pairwise interactions,
\begin{equation}
\label{eqn:airebo}
E=\frac{1}{2} \sum_i \sum_{j\neq i} \left[E_{ij}^{REBO} +E_{ij}^{LJ}+ \sum_{k\neq i,j} \sum_{l\neq i,j,k}  E_{kijl}^{TORS} \right] \;,
\end{equation} 
where $E_{ij}^{REBO}$ represents the covalent bonding interactions, $E_{kijl}^{TORS}$ is the torsion term that ensures the correct dihedral angles, and $E_{ij}^{LJ}$ is a modified Lennard-Jones term  accounting for the NCIs between the atoms \cite{stuart2000reactive}.  $E_{ij}^{LJ}$ only acts between atoms if they are not connected directly or indirectly through covalent bonds within a range between $2$ \AA \  and $8.5$ \AA. The AIREBO potential is  widely used for simulating molecules with more than $10^{3}$ carbon atoms, where \textit{ab initio} simulations are computationally too expensive.  

The initial building block for our DGM is a quadratic graphene membrane ( flat  hexagonal carbon mantle) with  an initial  bond length of 1.4 \AA\ (  which is close to the equilibrium bond-length of graphene)  and an edge length of $L= 60$ \AA.   The exact value of the initial bond-length is irrelevant, since the potential dynamically changes it,  and finally adjusts it to the equilibrium length.  To introduce disorder, we  create vacancies by randomly removing carbon atoms, obtaining vacancy concentrations ranging from $p=0$ to the critical percolation point $p_c=0.303$ for hexagonal lattices, which is the highest possible vacancy concentration  that can be achieved for our purpose, because only small clusters remain for higher values of $p$. After damaging the graphene membranes with vacancies, we extract the largest connected cluster.  Note that there is experimental evidence showing that if similar graphene clusters (or flakes) have a certain size, then fullerenes cages can be formed. \cite{chuvilin2010direct}  This carbon cluster is chemically very reactive because some carbon atoms are left with less than three neighbours.  We then reduce the reactivity of the system by adding hydrogen atoms. To each carbon atom with only two neighbours, we add a  hydrogen atom in  $z$-direction (randomly up or down in order to avoid a preferred crumpling direction). For carbon atoms with one single carbon neighbour, we add two hydrogen atoms, one in positive and one in negative  $z$-direction. 
  Note that adding hydrogen atoms is crucial for the stability of the DGM. If we do not add  hydrogen atoms to avoid the passivation of the dangling bonds, then we can only simulate vacancy concentrations up to 10\%. For higher vacancy probabilities, the graphene sheet cannot recover and gets more akin to amorphous carbon. Thus, the presence of hydrogen atoms inhibit the saturation of  
benzene rings in the disordered hexagonal lattice, allowing to keep $sp^2$ hybridisation and the conjugated bonds. 
By applying  this procedure, we obtain for $p=0$ a graphene sheet where the open edges with dangling $\sigma-$orbitals are terminated by carbon-hydrogen bonds. 
The densities of these DGMs exhibit a power-law dependence on $p$. 

We perform simulations with several realizations of DGMs for different vacancy densities $p$. 
We set the time step for the molecular mechanics simulation to $0.1$ fs, which is sufficiently small to capture the carbon-carbon and the carbon-hydrogen interaction properly. In order to fix the temperature, we apply a Nos\'{e}-Hoover thermostat obtaining the equilibrium state of the DGM  in the canonical ensemble (NVT). Each simulation begins with a flat DGM located in the $x-y$ plane at a temperature of $800$ K  which corresponds to an optimal value that is low enough  to keep covalent bonds, provides  enough kinetic energy to explore the phase space, and speeds up the equilibration process.   (see Fig. \ref{fig:initial_final_p69}a).  Subsequently, we cool down the structure gradually using a step size of $25$ K, giving enough time for equilibration at each temperature (see Fig. \ref{fig:initial_final_p69}b). To increase the precision during the last $25$ K, we decrease the temperature step size to $5$ K until we reach $0$~K  (see Fig.~\ref{fig:initial_final_p69}c  and section "computational details" in the supplementary material ). The equilibration time for each temperature step is chosen proportional to the number of carbon atoms contained in the DGM. Once the sheet reaches $0$~K, we start performing the measurements for the fractal dimension. Finally, in order to analyse the influence of the NCI, we deactivate the $E_{ij}^{LJ}$ interactions of the potential described in Eq.~\eqref{eqn:airebo}, repeat all simulations, and compare the results with the corresponding ones obtained when NCIs are present.

The obtained DGM structures are characterized in terms of the gyration tensor and the fractal dimension. We first compute the center of mass, i.e.   ${\vec{r}_{cm}}={1}/{M}\sum_{i=1}^N m_i \vec{r}_i$, where $M$ is the total mass of the structure (the mass of carbon atoms is set to 12 atomic units and the one of hydrogen atoms to 1 atomic unit), and shift the origin of the coordinate system to the center of mass frame. The gyration tensor is then obtained by 
\begin{equation}
\label{Smn}
S_{mn}= \frac{1}{M} \sum_{i=1}^N   m_i \left ({r}_i^ {(m)}-{r}_{cm}^ {(m)} \right) \left({r}_i^ {(n)}-{r}_{cm}^ {(n)}\right) \quad ,
\end{equation}
where $m,n\in \lbrace1,2,3\rbrace$. This matrix is symmetric and, therefore, its eigenvalues $\lambda_n$ are real and their associated  eigenvectors  orthogonal to each other. The eigenvalues $\lambda_n$ correspond to the extensions  in the direction of their eigenvectors. The sum of the eigenvalues gives the square of the radius of gyration,  $R_g^2=\lambda_1+\lambda_2+\lambda_3$. 
The radius of gyration $R_g$ is an appropriate measure to quantify the compactness of various structures, like for instance organic molecules \cite{RG_indicator_compactness}. 
Furthermore, if the structure is fractal, $R_g$ should relate with the mass $M$ of the total DGM  as a power law, 
\begin{equation}
R_g \sim  M^{\frac{1}{d_F}} \quad .
\label{Rg_M}
\end{equation}
To obtain the fractal dimension we use the sand-box method through the relation $M(r) \propto r^{d_F}$, where $M(r)$ is the mass of the atoms contained in a sphere of radius $r$ and origin at the closest atom to the center of mass  (See Fig.~S1 
of the Supplementary Material). We compute $M(r)$ in discrete exponential intervals, $r_k=1.05^k$, where $k\in \mathbb{N}$. 
With the sand-box method we obtain a $d_F$ for each single DGM.

\section*{Results and Discussion}

\begin{figure}
\begin{center}
\includegraphics[width=0.5\columnwidth]{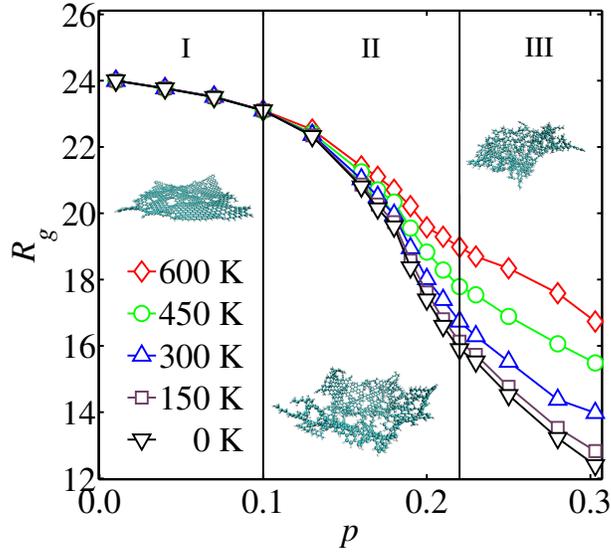}
\caption{The dependence of $R_g$ on the vacancy density $p$ for different  temperatures considering the presence of  NCIs. The snapshots of the DGMs were taken at $T=0$ K for $p=\lbrace 0.05,0.16,0.28\rbrace$. The error bars are smaller than the symbols.}
\label{fig:Rg_temperatures}
\end{center}
\end{figure}

Figure~\ref{fig:Rg_temperatures} shows the radius of gyration $R_g$ for different temperatures and vacancy densities $p$. We observe that $R_g$ is strongly dependent on  temperature  and can vary as much as $30\%$ for high values of $p$. This is in agreement with the fact that thermal fluctuations act stronger on the out-of-plane bending modes than on the in-plane stretching modes \cite{Xu2010}. After equilibration at $0$~K, where thermal fluctuations are absent, we obtain the most compact structure, i.e. having the smallest $R_g$, and find, by considering different system sizes, that DGMs  display self-similarity for all vacancy densities $p$ (see Supplementary Material). Furthermore, we see that $R_g$  decreases by increasing $p$. There are two explanations for this behaviour. First, the higher the value of $p$, the less carbon atoms are contained in the graphene membrane, leading to a less extended system. Second, and more important, the DGM undergoes a transition in region II ($0.1 < p < 0.22$, see Fig.~\ref{fig:Rg_temperatures}), leading to a more compact object, and consequently to a smaller value of  $R_g$.

\begin{figure}
\begin{center}
\includegraphics [width=0.5\columnwidth]{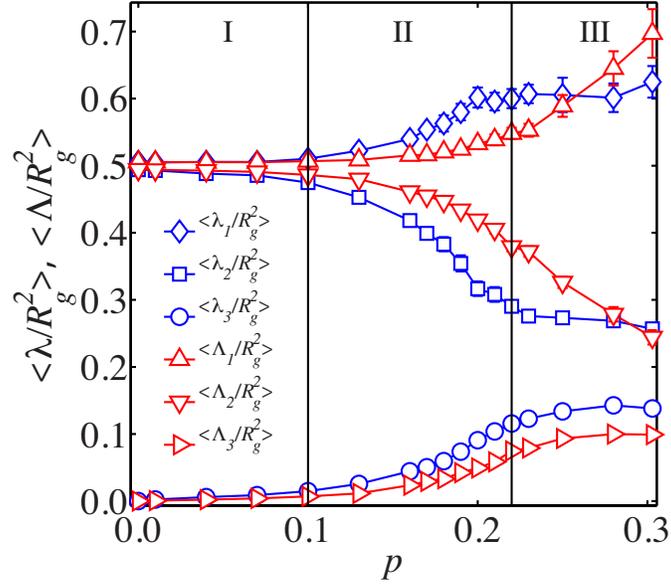}
\caption{
The dependence of the  normalised eigenvalues ${\lambda_n}/{R_g^2}$ and ${\Lambda_n}/{R_g^2}$  (for $n\in \lbrace1,2,3\rbrace$) of the gyration tensor on the vacancy density $p$ for simulations with all interactions (blue squares, circles and diamonds) and without NCIs (red triangles), respectively. }
\label{fig:Eval_gyration_tensor}
\end{center}
\end{figure}

\begin{figure}
\begin{center}
\includegraphics [width=0.5\columnwidth]{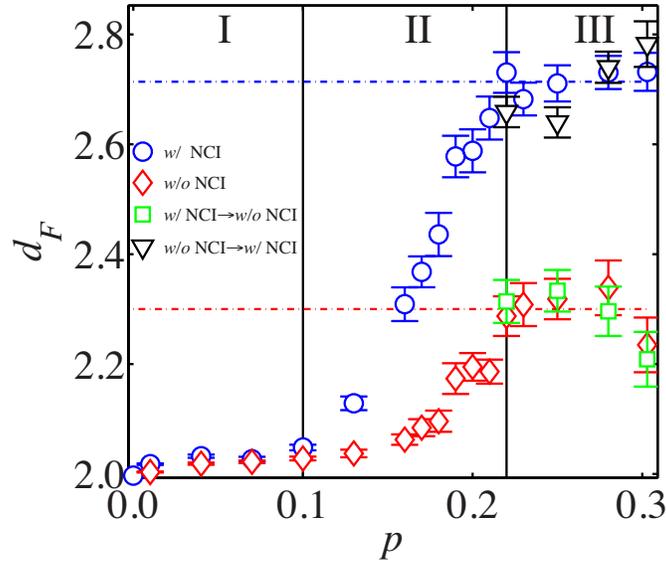}
\caption{Fractal dimension $d_F$  computed with the sand-box method for different fractions of vacancies $p$. 
The values of $d_F$ correspond to averages over 40 realizations of DGMs for $p<0.1$ and 100 realizations for all other values of $p$.
The blue circles correspond to the DGM after equilibration  considering all the interactions. The dashed blue line indicates the average of $d_F$ calculated over the values within  region III. 
The red diamonds correspond to $d_F$ after equilibration  in absence of NCIs. The dashed red line is the average of $d_F$ without NCIs for region III. 
The green squares result from simulations that were first performed with all interactions and, after that, equilibrated again without NCI.   
The black triangles correspond to  simulations that were first performed without NCIs and, after that, equilibrated again   considering all interactions.}
\label{fig:FD_SB}
\end{center}
\end{figure}

One should note that  $R_g$ does not provide information on how the DGM extends in different directions. 
This information can be retrieved from the normalised eigenvalues of the gyration tensor. We denote the eigenvalues of the gyration tensor of the DGM with NCIs by ${\lambda_n}/{R_g^2}$ and  the ones without NCIs by ${\Lambda_n}/{R_g^2} $
 (see Fig.~\ref{fig:Eval_gyration_tensor}).  
In region I ($p\leq0.1$), the DGM has only a few local vacancies and hydrogen atoms. The fractal dimension of both simulations (with and without NCIs) is $d_F \approx 2$ (See Fig.~\ref{fig:FD_SB}), indicating that the DGM remains virtually   a flat object  (exhibiting small ripples).  This finding is confirmed in Fig.~\ref{fig:Eval_gyration_tensor} by the fact that two eigenvalues have almost the same value and the third one is close to zero (${\lambda_3}/{R_g^2},{\Lambda_3}/{R_g^2}\approx 0$).

In contrast to region I, we observe that in region III the DGM extends in all three principal axes (${\lambda_3}/{R_g^2},{\Lambda_3}/{R_g^2} > 0$)  being  essentially isotropic close to the center of mass (within  a radius $\leq \min\lbrace\lambda_1,\lambda_2, \lambda_3\rbrace$, see also Supplementary Material Fig.~S3). 
In this case, the sheets crumple resulting in a  fractal dimension $d_F=2.71 \pm 0.02$, when all interactions between two atoms are considered. The three eigenvalues ($\lambda_n$) do not change significantly in this region. Surprisingly, the simulations without NCI reveal  a much smaller  fractal dimension, $d_F=2.30 \pm 0.05$, showing   that the NCI play a crucial role in compressing the DGMs. 
In principle, our simulation allows for the creation of new covalent bonds during the crumpling process. However, as shown in the Supplementary Material,  and contrary to the significant influence of the NCIs, the impact of newly formed covalent bonds on the crumpled structure is practically negligible. 

As depicted in Fig.~\ref{fig:FD_SB}, the crumpling transition takes place in the intermediate region II, for which  $0.1< p < 0.22$. 
Close to $p=0.1$, some stronger deformations orthogonal to the original plane of the hydrocarbon sheet ($p=0$) become visible (as an increase in ${\lambda_3}/{R_g^2},{\Lambda_3}/{R_g^2}$) 
and a characteristic direction for each DGM can be observed,
as reflected by an increase of the ratios ${\lambda_1}/{\lambda_2},{\Lambda_1}/{\Lambda_2}$. 
Note that the anisotropy reflected in the eigenvalues is a consequence of computing the eigenvalues for each single DGM instead of considering the average over all DGM samples together. 
Due to randomness, the underlying cluster from which each single DGM is constructed  has a characteristic direction and therefore $\lambda_1  > \lambda_2$.
This  anisotropy vanishes if we would overlap all DGMs at the center of mass and evaluate the eigenvalues of this structure.

 For low vacancy densities $p$, the  fractal dimension must be evaluated carefully due to the high anisotropy of the structure (see Supplementary Material). 
Therefore, the continuous change in $d_F$ in region II seems to be only a finite-size effect and thus we expect that this region will shrink to a transition point by increasing the system size, leading to  a discontinuous change in fractal dimension  in the thermodynamic limit.   This transition point is the value which separates the two phases: For low values of $p$ we have the flat phase (including some ripples and wrinkles) and for high values of $p$ we have the crumpled phase.

Note that from Fig.~\ref{fig:FD_SB}, we can also make some observations concerning the reversibility of switching on/off the NCIs. For that purpose, we performed simulations including all interactions and, afterwards, equilibrated again without NCIs (and vice versa). 
We found that the fractal dimension after equilibration doesn't depend on how the DGM was equilibrated before and therefore we can conclude that the process of switching on/off the NCIs is completely reversible.

Interestingly, fractal dimensions  comparable to our findings  have been observed experimentally for dry and fresh cream layers, $2.65 \pm 0.10$ and $2.45 \pm 0.15$, respectively \cite{Gomes2007}, as well as theoretically for saturated hydrocarbon structures at the critical point of percolation, $2.63\pm 0.15$ \cite{SoaresAndrade2005}.   
In the first case, the fractal dimension is higher for dry layers because the water molecules between the polymerised membrane is evaporated. The interactions with water molecules seem to screen the  NCIs within atoms of the polymerised membrane and, in analogy to our findings, produce a less compact structure.    
 A well defined fractal dimension is also observed in many other crumpling processes such as paper sheets \cite{GomesPaperFolding} and wires crumpled to spherical compact balls \cite{aguiar1991geometrical,PhysRevLett.101.094101}, just to name a few \cite{Balankin2007,PhysRevLett.58.2774}.
In Ref.~\cite{Debierre}, the authors studied self-avoiding surfaces as possible models of rapidly polymerising polymer membranes and found that self-avoiding  surfaces might act similarly to DGM without NCI, where only the short-range covalent-bond repulsion is left. As a consequence, the reported Monte Carlo simulations produced   self-avoiding surfaces with $d_F=2.35 \pm 0.05$, which is in agreement with our results. 
The crumpling driven by mechanically compressing an isotropic elastic shell that contains a  graphene membrane led to the same fractal dimension within error bars \cite{Buehler2011}. 

One can develop an effective theory to explain the fractal dimensions obtained in our simulations. For instance, a mean-field phenomenological model using an entropic elastic energy $U_S=AR^2$ \cite{keten2008strength}, and a two-body repulsive energy $U_{SA}=B\rho^ 2V$ leads to a mass-size scaling  $M\propto R^{2.5}$, which is consistent with our results  for DGM in absence of NCIs. NCIs expressed through a modified Lennard-Jones potential influence the repulsive energy $U_{SA}$ leading to a term $U_{SA}^{NCI}=C\rho^4V=C{M^4}/{R^9}$. After minimisation of $E=U_S+U_{SA}^{NCI}$ with respect to $R$, the mass-size relation $M \propto R^{2.75}$ follows straightforwardly, which is consistent with our results (see Supplementary Material for more details).

We expect an experimental  confirmation of our results to be, in principle, realizable  by applying two concepts: in the first one,  
the creation of vacancies can be obtained by  electron irradiation\cite{Kotakoski2011}, ion irradiation \cite{Lehtinen2011,Wang2012}, or by an adequate treatment with   plasmas; and in the second, the addition  of hydrogen atoms could be in principle performed by exposing the membrane to a cold hydrogen plasma \cite{Elias30012009}. 
For instance, one could think of a set-up, where sufficient hydrogen atoms are present during the damaging process such that recombination of carbon atoms does not take place and the removed carbon atoms gets directly replaced by hydrogen ones. This has been already observed for oxygen atoms which bind on sub-nanometer vacancy defects in the basal plane of graphene \cite{ja4117268}. We could expect a possible extension for hydrogen atoms. Additonally, one can damage graphane membranes instead of graphene \cite{PhysRevB.75.153401}, with an e-beam, and then using its reversible properties to release the residual hydrogen atoms that characterize graphane \cite{Elias30012009}.

\section*{Conclusion}

In summary, we have studied the crumpling transition of DGMs obtained by introducing random vacancies and hydrogen atoms to a graphene sheet.  We have shown that there is a clear transition from a flat membrane towards a crumpled DGM by increasing the fraction $p$ of vacancies. The crumpling transition has been analysed in terms of three different tools, namely,  the radius of gyration $R_g$, the eigenvalues of the gyration tensor, and the fractal dimension. We  observed that $R_g$ decreases by decreasing the temperature and by increasing   $p$, providing information on the way a DGM crumples. 
We could delimit three characteristic regions, depending on the degree of damage imposed to the graphene sheets:    in region I ($p<0.1$), the DGM is only extended in two principal axes and $d_F\approx 2$; in region II ($0.1<p<0.22$), a  system size dependent   transition occurs from   a flat  to a  crumpled object ; and in region III ($p>0.22$), we observe essentially isotropic and crumpled DGM. The crumpled graphene sheets are self-similar, with  a fractal dimension of $2.71 \pm 0.02$ and $2.30 \pm 0.05$ for simulations with and without NCIs, respectively. From this last result, we deduce that the NCI play a crucial role in the crumpling process during compression of DGMs. Finally, we also provide a phenomenological  model that describes  qualitatively our numerical findings.

\section*{Acknowledgements}
We thank the Brazilian agencies CNPq, CAPES, and
FUNCAP, the National Institute of Science and Technology for
Complex Systems in Brazil, and the European Research
Council (ERC) Advanced Grant No. 319968-FlowCCS
for financial support.

\section*{Author Contributions}
H.H. and J.A. provided the idea and the methods used in the paper. 
M.G. developed a theoretical model to describe the fractal dimension. 
I.G. performed the simulations. 
M.M. and I.G. processed the data and applied the methods described in the paper.
J.A. and H.H. contributed to the analysis of the data.  
H.H. and M.M. supervised the work. 
All authors contributed to the writing process and reviewed the manuscript.

\section*{Additional Information}
Supplementary information accompanies this paper at http://www.nature.com/srep.
\newline
\textbf{Competing financial interests:} The authors declare no competing financial interests.

%
%
%
%
%
\end{document}